\documentclass[conference]{IEEEtran}
\IEEEoverridecommandlockouts
\usepackage{cite}
\usepackage{amsmath,amssymb,amsfonts}
\allowdisplaybreaks
\usepackage{bm}
\usepackage{graphicx}
\usepackage{textcomp}
\usepackage{xcolor}
\usepackage[
    colorlinks=true,
    citecolor=blue,
    linkcolor=.,
    urlcolor=blue
]{hyperref}
\usepackage{glossaries}
\newacronym{ac}{AC}{alternating current}
\newacronym{ad}{AD}{automatic differentiation}
\newacronym{dc}{DC}{direct current}
\newacronym{opf}{OPF}{optimal power flow}
\newacronym{tso}{TSO}{transmission system operators}
\newacronym{gpu}{GPU}{graphics processing unit}
\newacronym{cpu}{CPU}{central processing unit}
\newacronym{sc}{SC}{security constrained}
\newacronym{soc}{SOC}{second-order cone}
\newacronym{qc}{QC}{quadratic convex}
\newacronym{agc}{AGC}{automatic generation control}
\newacronym{mafr}{MAFR}{minimum-adjustment feasibility restoration}
\def\BibTeX{{\rm B\kern-.05em{\sc i\kern-.025em b}\kern-.08em
    T\kern-.1667em\lower.7ex\hbox{E}\kern-.125emX}}
\usepackage[nameinlink]{cleveref}

\usepackage{algorithm}
\usepackage{algpseudocode}
\newtheorem{remark}{Remark}

\begin{document}
\title{Quantifying the Cost of Restoring AC Feasibility for Approximate Optimal Power Flow Solutions\\
\thanks{This material is based upon work supported by the National Science Foundation Graduate Research Fellowship under Grant No. DGE-2141064.}
}

\author{\IEEEauthorblockN{Sanjay Johnson, Sungho Shin}
\IEEEauthorblockA{Massachusetts Institute of Technology \\
Cambridge, MA, USA \\
\{sanjayjo, sushin\}@mit.edu}
}

\maketitle

\begin{abstract}
Power system operations rely on approximate solutions of the AC optimal power flow problem, most notably the DC optimal power flow used in security-constrained unit commitment and economic dispatch, while convex relaxations, machine learning surrogates, and other approximations are studied extensively in the academic literature. Because these approximations neglect or relax parts of the AC power flow equations, their solutions must be adjusted before dispatch; for example, the DC optimal power flow, which neglects transmission losses, reactive power, and voltage magnitude variations, is known to produce solutions that are always infeasible. The cost of this adjustment is not adequately understood. This paper proposes a method that quantifies, for any approximate solution, the minimum cost of the adjustments that restore AC feasibility, which is a lower bound on the cost of the heuristic adjustments used in practice. We apply it to the DC optimal power flow, the second-order cone and quadratic convex relaxations, and the AC optimal power flow solved to a loose relative tolerance of $10^{-4}$ instead of $10^{-8}$. Relative to the AC optimal power flow solution obtained from a local nonlinear optimization solver, and even with the minimum adjustment, the DC optimal power flow requires adjusting, on average, generators that account for 15.6\% of total generation capacity and incurs a 1.10\% cost increase; the convex relaxations require more adjustment (17.2\% and 16.9\%) but smaller cost increases (0.68\% and 0.78\%); and the loose-tolerance AC optimal power flow requires 6.1\% and incurs 0.42\%. These results indicate significant value in operating markets directly on the AC optimal power flow, which avoids the distortion of the dispatch and prices caused by heuristic feasibility restoration.
\end{abstract}


\section{Introduction}
\label{introduction}

In practice, \gls{tso} operate markets through security-constrained unit commitment and economic dispatch, both of which rely on a linearized model known as the \gls{dc} \gls{opf} \cite{oneill_recent_2011, chen_security-constrained_2023}. The \gls{dc} \gls{opf} is a convex quadratic program and is therefore fast to solve, but its solutions cannot be delivered on the physical grid as computed and must be adjusted before dispatch \cite{baker_solutions_2021}. The linearization neglects transmission losses, reactive power, and the other nonlinear relations that govern the \gls{ac} system. Operators therefore restore feasibility of the \gls{ac} power flow equations through a heuristic iterative process. The resulting dispatch is in general suboptimal for the \gls{ac} system, but the cost of adjustments based on such heuristics is not quantitatively understood.

\Gls{ac} \gls*{opf} models instead find operating points that are optimal for the \gls{ac} system, considering both efficient economic dispatch and mitigation of potential security contingencies ~\cite{cain_history_2012}. Because active power, reactive power, and the nonlinear \gls{ac} power flow equations are considered simultaneously, the \gls{ac} \gls{opf} removes the suboptimality caused by the heuristic iterative process for restoring \gls{ac} feasibility.

The \gls{ac} \gls{opf} has been avoided in practice mostly because of its computational challenges, which are of two kinds. The first is solving the \gls{ac} power flow equations and unit commitment simultaneously, where the nonlinear power flow equations are coupled with discrete commitment decisions and the problem becomes a mixed-integer nonlinear program. This remains an open question, and was the focus of the recent Grid Optimization Competition \cite{holzer_go_2024}. The second is solving the \gls{ac} \gls{opf} itself, even with the unit commitment fixed: the nonlinear, nonconvex power flow equations must be satisfied over networks with thousands of buses within operational timescales. This is no longer a major obstacle, since mature local solvers such as IPOPT \cite{wachter_implementation_2006} and KNITRO \cite{byrd_knitro_2006} exist, and \gls{gpu}-based solvers such as MadNLP \cite{shin_accelerating_2024} now accelerate them further. For example, an \gls{ac} \gls{opf} with 78,484 buses is solved in about 16~s on a \gls{gpu}, compared with about 390~s on a CPU \cite{johnson_examodelspowerjl_2026}. 

The practical implications of using \gls{dc} \gls{opf} solutions along with the heuristic adjustment procedure, as opposed to directly solving the \gls{ac} \gls{opf}, have not been adequately quantified and reported in the literature. This paper aims to address the following research question:
\begin{quote}
\itshape What is the practical cost of dispatching an approximate \gls{opf} solution after heuristic adjustment, instead of the solution of the \gls{ac} \gls{opf}?
\end{quote}

\paragraph*{Our Approach} We address this question by proposing a method, which we call \gls{mafr}, that determines the minimum number of generators whose active power must change from an approximate solution to restore \gls{ac} feasibility, counting generators regardless of the size of their adjustment, and the least cost achievable with that adjustment. We also explore the further cost reduction resulting from each subsequent adjustment that is made after feasibility is achieved. This gives the minimum additional cost incurred by using an approximate solution method instead of the exactly solved \gls{ac} \gls{opf}, for any given degree of intervention by a \gls{tso}, and hence a lower bound on the cost of the heuristic adjustments used in practice. Our numerical studies consider the \gls{dc} \gls{opf}, the \gls{soc} and \gls{qc} relaxations, and the \gls{ac} \gls{opf} solved to a loose relative tolerance of $10^{-4}$ instead of $10^{-8}$, which we call the loose-tolerance \gls{ac} \gls{opf}, as candidate methods, and analyze all four on equal footing.

\paragraph*{Related Work} Baker \cite{baker_solutions_2021} shows that, under mild assumptions, no solution of the \gls{dc} \gls{opf} satisfies the \gls{ac} power flow equations, even when the \gls{dc} dispatch is adjusted to account for losses. Boateng et al. \cite{boateng_towards_2027} develop a pipeline that recovers \gls{ac} feasible solutions from dispatches produced by four \gls{dc} \gls{opf} variants, using distributed slack allocation and PV/PQ switching, and evaluate it over more than 10,000 dispatch scenarios. These works show that \gls{dc} \gls{opf} dispatches must be corrected and how to correct them. In contrast, our work does not propose a correction procedure; it identifies the minimum adjustment needed to restore \gls{ac} feasibility and the resulting cost, which serves as a best-case reference for any procedure which restricts adjustments to select generators.

\paragraph*{Contributions} Our contributions are as follows.
\begin{enumerate}
\item We introduce \gls{mafr} (\Cref{methodology}). To the best of our knowledge, it is the first method that quantifies the minimum adjustment required to restore \gls{ac} feasibility, and its cost, independently of a specific correction procedure and for any approximate solution method.
\item We apply \gls{mafr} to the \gls{dc} \gls{opf}, the \gls{soc} and \gls{qc} relaxations, and the loose-tolerance \gls{ac} \gls{opf} on the PGLib-OPF cases \cite{babaeinejadsarookolaee2019power}, and demonstrate that, even with the minimum adjustment, the \gls{dc} \gls{opf} requires adjusting generators that account for 15.6\% of total generation capacity and incurs a 1.10\% cost increase on average, the relaxations require more adjustment but smaller cost increases, and the loose-tolerance \gls{ac} \gls{opf} requires the least of both. To the best of our knowledge, this is the first comparison of the adjustment cost across approximate solution methods.
\end{enumerate}

\section{Power Flow Formulations}
\label{formulations}

\subsection{Baseline AC Optimal Power Flow Model}
\label{baseline}

The baseline \gls{ac} \gls{opf}, the basis for our comparison, is

\begin{subequations}
\label{eq:ac}
\begin{align}
    \min_{\substack{S^g,\, V,\\ S,\, \alpha}} \quad
    & J(S^g,\alpha) := \sum_{k \in \mathcal{G}} \big( c_{2k} (\Re S^g_k)^2 + c_{1k} \Re S^g_k + c_{0k} \big) \nonumber \\
    & \hspace{6em} + c_{\mathrm{shed}} \sum_{i \in \mathcal{N}} \alpha_i | S^d_i |
    \label{ac_obj} \\
    \text{s.t.} \quad
    & \angle V_{r} = 0
    \label{ac_con1} \\
    & S^{gl}_k \leq S^g_k \leq S^{gu}_k
    \quad \forall k \in \mathcal{G}
    \label{ac_con2} \\
    & v^l_i \leq |V_i| \leq v^u_i,\ 0 \leq \alpha_i \leq 1 \quad \forall i \in \mathcal{N}
    \label{ac_con3} \\
    & \sum_{k \in \mathcal{G}_i} S^g_k - (1-\alpha_i) S^d_i - Y^s_i |V_i|^2 = \sum_{(i,j) \in \mathcal{E}_i} S_{ij}
    \label{ac_con4} \\
    & S_{ij} = \left( Y^*_{ij} - \mathrm{i} \frac{b^c_{ij}}{2} \right)
      |V_i / T_{ij}|^2
      - Y^*_{ij} V_i V^*_j / T_{ij}
    \label{ac_con5} \\
    & S_{ji} = \left( Y^*_{ij} - \mathrm{i} \frac{b^c_{ij}}{2} \right) |V_j|^2
      - Y^*_{ij} V^*_i V_j / T^*_{ij}
    \label{ac_con6} \\
    & |S_{ij}| \leq s^u_{ij}
    \quad \forall (i,j) \in \mathcal{E} \cup \mathcal{E}^R
    \label{ac_con7} \\
    & \theta^{\Delta l}_{ij} \leq \angle (V_i V^*_j) \leq \theta^{\Delta u}_{ij}
    \quad \forall (i,j) \in \mathcal{E}
    \label{ac_con8}
\end{align}
\end{subequations}
where $\mathcal{G}$, $\mathcal{N}$, and $\mathcal{E}$ are the sets of generators, buses, and branches; $\mathcal{G}_i \subseteq \mathcal{G}$ is the set of generators at bus $i$; $\mathcal{E}^R$ is the set of branches in reverse orientation, and $\mathcal{E}_i$ is the set of branches in $\mathcal{E} \cup \mathcal{E}^R$ originating at bus $i$; $r$ is the reference bus; $S^g_k \in \mathbb{C}$ is the complex power generation of generator $k$, $c_{2k}, c_{1k}, c_{0k} \in \mathbb{R}$ are its cost coefficients, and $S^{gl}_k, S^{gu}_k \in \mathbb{C}$ are its power limits; $V_i \in \mathbb{C}$ is the complex voltage at bus $i$, and $v^l_i, v^u_i \in \mathbb{R}$ are its magnitude limits; $S^d_i \in \mathbb{C}$ is the complex power demand, $Y^s_i \in \mathbb{C}$ the shunt admittance, and $\alpha_i \in [0,1]$ the fraction of the demand that is shed at bus $i$; $S_{ij} \in \mathbb{C}$ is the complex power flow on branch $(i,j)$, $Y_{ij} \in \mathbb{C}$ is the branch admittance, $b^c_{ij} \in \mathbb{R}$ is the line charging susceptance, $T_{ij} \in \mathbb{C}$ is the transformer tap ratio, $s^u_{ij} \in \mathbb{R}$ is the apparent power limit, and $\theta^{\Delta l}_{ij}, \theta^{\Delta u}_{ij} \in \mathbb{R}$ are the voltage angle-difference limits; and $\Re(\cdot)$, $(\cdot)^*$, and $\angle(\cdot)$ denote the real part, the complex conjugate, and the phase angle. The inequalities between complex quantities in \eqref{ac_con2} hold for the real and imaginary parts separately. The complex quantities are written in terms of their real and imaginary parts as $S^g_k = P^g_k + \mathrm{i} Q^g_k$, $S^d_i = P^d_i + \mathrm{i} Q^d_i$, $S_{ij} = P_{ij} + \mathrm{i} Q_{ij}$, $V_i = |V_i| e^{\mathrm{i}\theta_i}$, $Y_{ij} = G_{ij} + \mathrm{i} B_{ij}$, and $Y^s_i = G^s_i + \mathrm{i} B^s_i$, where $\mathrm{i} = \sqrt{-1}$ is the imaginary unit, $P, Q \in \mathbb{R}$ denote active and reactive power, $|V_i|, \theta_i \in \mathbb{R}$ are the voltage magnitude and angle at bus $i$, and $G, B \in \mathbb{R}$ denote conductance and susceptance.

The objective $J$ in \eqref{ac_obj} is the total active power generation cost plus the load shedding cost, where $c_{\mathrm{shed}} > 0$ is the cost per unit of shed demand. Constraint \eqref{ac_con1} fixes the voltage angle at the reference bus, \eqref{ac_con2} and \eqref{ac_con3} impose the generator power limits, the voltage magnitude limits, and the bounds on the fraction of demand that is shed, \eqref{ac_con4} enforces complex power balance at each bus $i \in \mathcal{N}$, \eqref{ac_con5} and \eqref{ac_con6}, imposed for all $(i,j) \in \mathcal{E}$, define the power flows in both directions of each branch, \eqref{ac_con7} limits the apparent power flow on each branch, and \eqref{ac_con8} limits the voltage angle difference across each branch.

Problem \eqref{eq:ac} can be written in polar or rectangular coordinates, depending on whether each complex voltage is represented by its magnitude and angle or by its real and imaginary parts. In either representation, the power flow constraints \eqref{ac_con4}--\eqref{ac_con6} are nonconvex, so \eqref{eq:ac} is a nonlinear, nonconvex optimization problem.

We allow load shedding in \eqref{eq:ac} so that the problem remains feasible when the dispatch is restricted, as in \Cref{methodology}. The cost $c_{\mathrm{shed}}$ is set much higher than any generator cost, playing the role of the value of lost load (VOLL) in market clearing, so that load is shed only when no feasible dispatch exists without it; and because $\alpha_i$ scales the complex demand, the shed load keeps the power factor of the original demand.

\subsection{Approximate Optimal Power Flow Models}
\label{approx}

Problem \eqref{eq:ac} is approximated by various other formulations. The \gls{dc} \gls{opf} is perhaps one of the most widely used; it is obtained from \eqref{eq:ac} by neglecting transmission losses, reactive power, and voltage magnitude variations. It minimizes the generation cost in \eqref{ac_obj}, written in terms of $P^g_k$, subject to
\begin{equation}
\label{eq:dc}
    \sum_{k \in \mathcal{G}_i} P^g_k - P^d_i - G^s_i = \sum_{(i,j) \in \mathcal{E}_i} P_{ij}, \quad P_{ij} = -b^c_{ij} (\theta_i - \theta_j)
\end{equation}
where $P^g_k \in \mathbb{R}$ is the active power generation, $P_{ij} \in \mathbb{R}$ is the active power flow on branch $(i,j)$, and $\theta_i \in \mathbb{R}$ is the voltage angle at bus $i$; the two equations in \eqref{eq:dc} hold for all $i \in \mathcal{N}$ and $(i,j) \in \mathcal{E}$, respectively, together with $\theta_r = 0$ and the limits \eqref{ac_con2}, \eqref{ac_con7}, and \eqref{ac_con8} written in terms of $P^g_k$, $P_{ij}$, and $\theta_i - \theta_j$.

Several convex relaxations of the \gls{ac} \gls{opf} have also been proposed, including the semidefinite programming relaxation \cite{lavaei_zero_2012}, the \gls{soc} relaxation \cite{yuan_second-order_2019}, and the \gls{qc} relaxation \cite{coffrin_qc_2016}. We consider the \gls{soc} and \gls{qc} relaxations.

\section{Minimum-Adjustment Feasibility Restoration}
\label{methodology}

This section presents \gls{mafr}, the main contribution of this paper. The \gls{dc} \gls{opf} \eqref{eq:dc} and the convex relaxations \cite{lavaei_zero_2012, yuan_second-order_2019, coffrin_qc_2016} introduced in \Cref{approx} are only approximate, so their solutions must be adjusted to restore feasibility with respect to \eqref{eq:ac} (\Cref{introduction}). We consider the setting in which the operator has a limited budget of adjustments, that is, may change the active power of only a limited number of generators, and seeks the least-cost restoration of feasibility with respect to \eqref{eq:ac} within that budget. \gls{mafr} finds this least-cost restoration for each budget, using three modified formulations of the \gls{ac} \gls{opf}. It therefore estimates the best possible suboptimality, that is, the smallest adjustment cost achievable within a given budget; the suboptimality of the heuristic procedures used in practice can be higher. 

\subsection{Exact MAFR}
The exact formulation of \gls{mafr} minimizes the number of generators whose active power deviates from an approximate solution, using binary variables and a big-M constant. We refer to it as the exact \gls{mafr} problem:
\begin{subequations}
\label{bin_opf}
\begin{align}
    \min_{\substack{S^g,\, V,\, S,\, \alpha\\ z_k \in \{0,1\}}} \quad
    & J(S^g,\alpha) + c_{\mathrm{shift}} \sum_{k \in \mathcal{G}} z_k
      \label{bin_obj} \\
    \text{s.t.} \quad
    & \text{\eqref{ac_con1}--\eqref{ac_con8}},\ | P^g_k - P^g_{k,\mathrm{sol}} | \leq Mz_k \ \forall k \in \mathcal{G}
        \label{shift_con}
\end{align}
\end{subequations}
where $c_{\mathrm{shift}} > 0$ is the cost of adjusting one generator, $z_k \in \{0,1\}$ indicates whether generator $k$ is adjusted, $M > 0$ is a sufficiently large constant, and $P^g_{k,\mathrm{sol}} \in \mathbb{R}$ is the active power of generator $k$ in the approximate solution. Constraint \eqref{shift_con} is a big-M formulation: $z_k = 0$ forces $P^g_k = P^g_{k,\mathrm{sol}}$, while $z_k = 1$ leaves $P^g_k$ free to deviate.

The cost $c_{\mathrm{shift}}$ is set high enough that minimizing the number of generation changes is prioritized before minimizing cost. The load shedding cost $c_{\mathrm{shed}}$ is set higher than $c_{\mathrm{shift}}$, so that load is shed only when no set of adjustments restores \gls{ac} feasibility.

Solved to global optimality, the exact \gls{mafr} problem \eqref{bin_opf} gives the minimum number of adjustments and, after removing $c_{\mathrm{shift}} \sum_k z_k$ from its objective value, the resulting cost, which answers the research question posed in \Cref{introduction}. However, it is a mixed-integer nonlinear program and is intractable for large systems, so we estimate its solution with the two nonlinear programs below. 

\subsection{Adjustment Set Identification}
\label{sec:asi}
The first nonlinear program replaces the count of adjusted generators in the exact \gls{mafr} problem \eqref{bin_opf} with an $\ell_1$ penalty, a standard convex surrogate that promotes sparse solutions \cite{tibshirani_regression_1996}. We refer to it as the relaxed \gls{mafr} problem:
\begin{equation}
\label{Lmod_opf}
    \min_{\substack{S^g,\, V,\\ S,\, \alpha}} \ J(S^g,\alpha) + \lambda \sum_{k \in \mathcal{G}} | P^g_k - P^g_{k,\mathrm{sol}} | \ \ \text{s.t.} \ \text{\eqref{ac_con1}--\eqref{ac_con8}}
\end{equation}
where $\lambda > 0$ is the penalty parameter.

The relaxed \gls{mafr} problem \eqref{Lmod_opf} is used to rank the generators by how much adjusting them reduces load shedding and cost. Starting from a large $\lambda$, we solve it repeatedly, halving $\lambda$ each time, and append the generators that deviate from $P^g_{k,\mathrm{sol}}$ to an ordered list, in descending order of $|P^g_k - P^g_{k,\mathrm{sol}}|$; the remaining generators are appended in descending order of their maximum active power output (\Cref{alg:generate_gl}). The result is a ranking of all generators, written as a bijection $\pi : \{1, \ldots, |\mathcal{G}|\} \to \mathcal{G}$, where $\pi(j)$ is the $j$-th generator in the ranking.

\subsection{Adjustment Cost Evaluation}
The second nonlinear program fixes the active power of every generator outside a flexible set and re-optimizes the rest, which estimates the dispatch after adjustment. We refer to it as the cost evaluation problem: for $m = 0, 1, \ldots, |\mathcal{G}|$,
\begin{equation}
\label{mod_opf}
    \min_{\substack{S^g,\, V,\\ S,\, \alpha}} \ J(S^g,\alpha) \ \ \text{s.t.} \ \text{\eqref{ac_con1}--\eqref{ac_con8}},\ P^g_k = P^g_{k,\mathrm{sol}} \ \forall k \notin \mathcal{F}_m
\end{equation}
where $\mathcal{F}_m = \pi(\{1, \ldots, m\})$ is the set of the first $m$ generators in the ranking $\pi$ from \Cref{sec:asi}, which we call the flexible set.

The two problems are not equivalent. The relaxed \gls{mafr} problem \eqref{Lmod_opf} is used only to rank the generators; it penalizes the deviation of every generator, including those in $\mathcal{F}_m$, which biases their setpoints toward $P^g_{k,\mathrm{sol}}$. The cost evaluation problem \eqref{mod_opf} removes this penalty and fixes the generators outside $\mathcal{F}_m$ exactly, so for each $\mathcal{F}_m$ it gives the least-cost adjustment of the generators in $\mathcal{F}_m$, free of the bias and of the small nonzero deviations that the slack-variable formulation of the $\ell_1$ term leaves in floating-point arithmetic. 

Because reactive power, voltage magnitudes, and voltage angles are also optimized, the objective value of the cost evaluation problem \eqref{mod_opf} is a lower bound on the cost of the dispatch after adjustment with $\mathcal{F}_m$.

Operators could in principle adjust the dispatch further once \gls{ac} feasibility is restored, in which case the objective at the smallest feasible $m$ overestimates the cost. Existing literature suggests that operators adjust only to address \gls{ac} nonlinearity \cite{noauthor_technical_2010, chen_security-constrained_2023}, so we assume no further adjustments. 

\Cref{alg:generate_gl} summarizes \gls{mafr}.

\begin{algorithm}[t]
\caption{Minimum-Adjustment Feasibility Restoration (MAFR) of a Single Approximate OPF Solution}
\label{alg:generate_gl}
\begin{algorithmic}[1]
\Require Case data; approximate dispatch $P^g_{k,\mathrm{sol}}$, clipped to the generator limits
\State $\pi \gets ()$, $\lambda \gets 10^8$
\Repeat \Comment{adjustment set identification}
    \State Solve the relaxed \gls{mafr} problem \eqref{Lmod_opf} with penalty $\lambda$
    \State Append newly adjusted generators to $\pi$, in descending order of $|P^g_k - P^g_{k,\mathrm{sol}}|$
    \State $\lambda \gets \lambda/2$
\Until{$\lambda$ is sufficiently small}
\State Append the remaining generators to $\pi$ in descending order of their maximum active power output
\For{each size $m$ on a grid from $0$ to $|\mathcal{G}|$} \Comment{adjustment cost evaluation}
    \State $\mathcal{F}_m \gets \pi(\{1, \ldots, m\})$
    \State Solve the cost evaluation problem \eqref{mod_opf}; record its objective and $\max_i \alpha_i$
\EndFor
\end{algorithmic}
\end{algorithm}

\begin{remark}
\gls{mafr} involves two approximations. First, the ranking $\pi$ is not guaranteed to be optimal, since determining the exact ranking would require a mixed-integer nonlinear program, which does not scale to the systems considered here. Second, the nonlinear programs in \gls{mafr} are solved to local optimality only, so the reported costs are those of locally optimal solutions. We assume that both approximations are sufficient for estimating the minimum adjustment and its cost.
\end{remark}

\section{Numerical Experiments}
\label{results}

\begin{figure*}[t]
  \centering
  \includegraphics[width=0.48\textwidth]{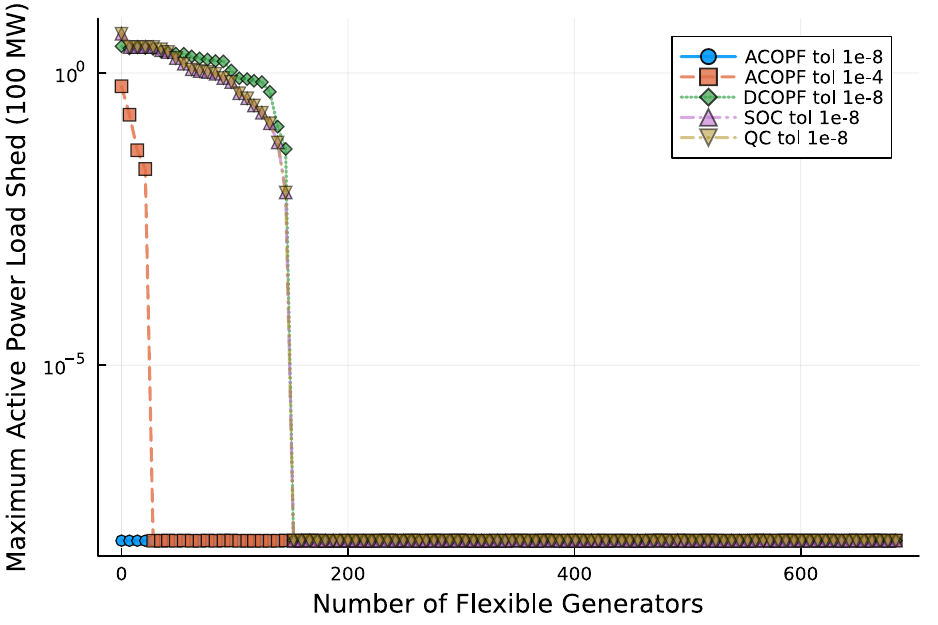}\hfill
  \includegraphics[width=0.48\textwidth]{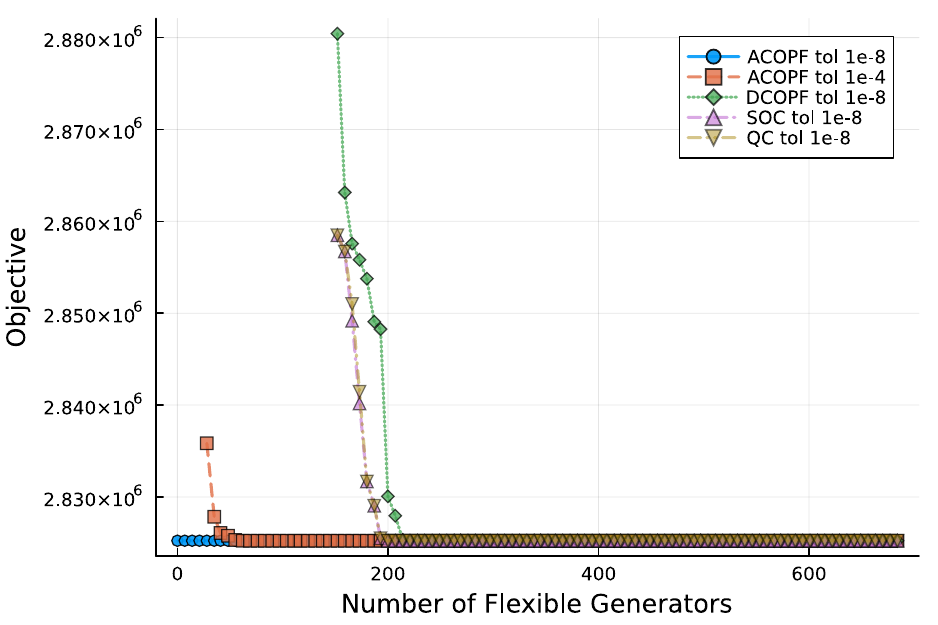}\\
  \includegraphics[width=0.48\textwidth]{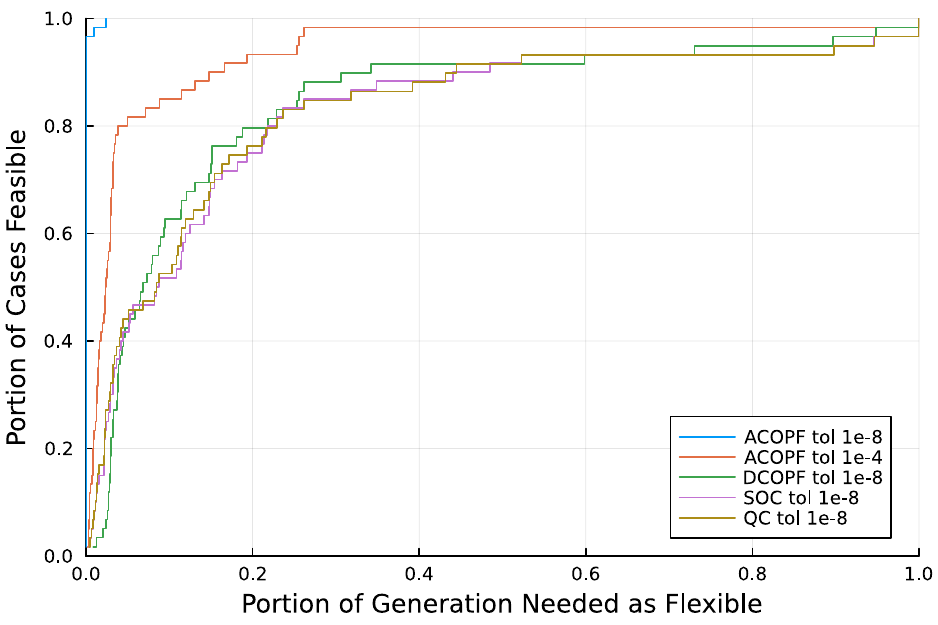}\hfill
  \includegraphics[width=0.48\textwidth]{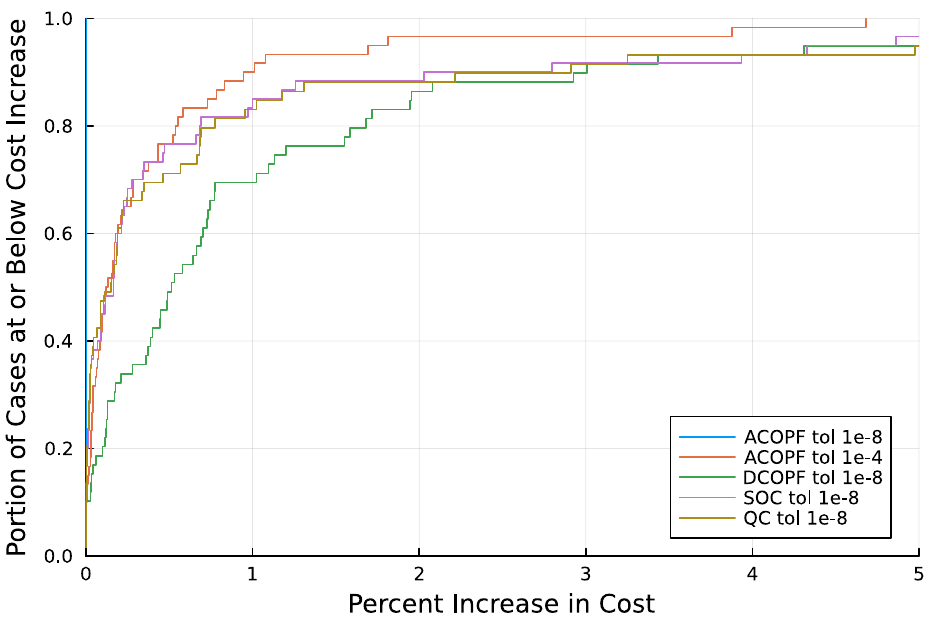}
  \caption{Top left: load shedding vs.\ number of flexible generators, pglib\_opf\_case6515\_rte. Top right: objective vs.\ number of flexible generators after AC feasibility is achieved, pglib\_opf\_case6515\_rte. Bottom left: CDF of the flexible capacity needed to reach AC feasibility, all cases. Bottom right: CDF of the cost increase at the smallest flexible set without load shedding, all cases.}
  \label{fig:results}
\end{figure*}

We model the \gls{ac} \gls{opf} and the \gls{dc} \gls{opf} in ExaModelsPower \cite{johnson_examodelspowerjl_2026} and the \gls{soc} and \gls{qc} relaxations in PowerModels \cite{coffrin_powermodels_2018}, and solve all problems to a relative tolerance of $10^{-8}$, except the loose-tolerance \gls{ac} \gls{opf} ($10^{-4}$). We apply \gls{mafr} to the solution of each method on all PGLib-OPF cases under typical operating conditions \cite{babaeinejadsarookolaee2019power}, excluding cases in which a method does not converge. We solve the cost evaluation problem \eqref{mod_opf} for $m$ on a grid of up to 101 values from $0$ to $|\mathcal{G}|$ (\Cref{alg:generate_gl}). The smallest $m$ with $\alpha = 0$ estimates the minimum number of adjustments needed to restore \gls{ac} feasibility, and the objective at that $m$ estimates the resulting cost, which we compare across methods and with the \gls{ac} \gls{opf} solution. The code is available at \url{https://github.com/mit-shin-group/power-flow-benchmark}. 

\Cref{fig:results} (top) shows, for pglib\_opf\_case6515\_rte, the load shedding (left) and, for points without load shedding, the objective of the cost evaluation problem \eqref{mod_opf} (right) against the number of flexible generators. 

For this case, the loose-tolerance \gls{ac} \gls{opf} needs only 28 flexible generators to reach \gls{ac} feasibility without load shedding, while the \gls{soc}, \gls{qc}, and \gls{dc} methods each need 152. At the earliest solution for each method requiring no load shedding, the loose-tolerance \gls{ac} \gls{opf} incurred a 0.38\% cost increase compared to the tight-tolerance \gls{ac} \gls{opf}, the \gls{soc} and \gls{qc} relaxations incurred a 1.18\% increase each, and the \gls{dc} \gls{opf} incurred a 1.95\% increase.

This ordering does not necessarily hold for every case; \Cref{fig:results} (bottom left and right) summarizes the results over all cases.

\Cref{fig:results} (bottom left) shows the fraction of cases that reach \gls{ac} feasibility as a function of the flexible capacity, defined as the capacity of $\mathcal{F}_m$ as a ratio of total generation capacity. The average flexible capacity needed is 6.1\%, 15.6\%, 17.2\%, and 16.9\% for the loose-tolerance \gls{ac} \gls{opf}, the \gls{dc} \gls{opf}, and the \gls{soc} and \gls{qc} relaxations, respectively. 

\Cref{fig:results} (bottom right) shows the distribution of the cost increase at the smallest flexible set without load shedding. The average cost increase is 0.42\%, 1.10\%, 0.68\%, and 0.78\%, respectively, and for every method a fraction of cases reaches \gls{ac} feasibility with no cost increase relative to the tight-tolerance \gls{ac} \gls{opf}.

\paragraph*{Discussion} The results point to significant value in solving the \gls{ac} \gls{opf} directly. Restoring \gls{ac} feasibility after the fact requires substantial changes to the dispatch, with generators accounting for 15.6\% of total capacity adjusted on average for the \gls{dc} \gls{opf} even with the minimum adjustment; such changes can distort the market mechanism, since the dispatched schedule then differs from the one on which prices are set, and they raise the cost, by 1.10\% on average. Operating the market directly on the \gls{ac} \gls{opf} would keep the locational marginal prices consistent with the dispatch and improve system-wide performance; even the loose-tolerance \gls{ac} \gls{opf}, which can be solved quickly on \glspl{gpu}, requires adjusting only 6.1\% of capacity at a 0.42\% cost increase. The convex relaxations require more adjustment than the \gls{dc} \gls{opf} (17.2\% and 16.9\% versus 15.6\%) but incur a smaller cost increase (0.68\% and 0.78\% versus 1.10\%), suggesting that modeling losses and reactive power places the dispatch closer to the \gls{ac} optimum in cost even when more generators must be moved.

\paragraph*{Limitations} Our analysis allows free adjustment of reactive power and voltage, assumes no further adjustments once \gls{ac} feasibility is restored, relies on a heuristic ranking and locally optimal solutions (Remark~1), and does not evaluate machine learning surrogates.

\section{Conclusion}
\label{conclusion}
This paper introduced \gls{mafr}, a method that estimates, for any approximate \gls{opf} solution, the minimum adjustment needed to restore \gls{ac} feasibility and the resulting cost. On the PGLib-OPF cases, even the minimum adjustment of a \gls{dc} \gls{opf} solution moves generators that account for 15.6\% of total generation capacity and raises the cost by 1.10\% on average; the convex relaxations require more adjustment at a smaller cost, and the loose-tolerance \gls{ac} \gls{opf} requires the least of both. These results indicate significant value in operating markets directly on the \gls{ac} \gls{opf}, which keeps prices consistent with the dispatch and reduces cost. Future work will identify the drivers of the differences between methods, extend \gls{mafr} to the security-constrained formulation, and provide it as a platform for evaluating new \gls{opf} methods.

\bibliographystyle{ieeetr} 
\bibliography{ExaPowerAnalysis}

\end{document}